\documentclass[10pt,
preprintnumbers,amsmath,amssymb,
aps,prd,nofootinbib,eqsecnum,a4paper]{revtex4}
\usepackage{graphicx,epsf}
\usepackage{xcolor}
\def\be{\begin{equation}}
\def\ee{\end{equation}}
\def\bea{\begin{eqnarray}}
\def\eea{\end{eqnarray}}
\def\cs2{c_{\rm{s}}^2}

\def\PY{{\sc{Pyessence}}}

\usepackage{float}

\newcommand{\C}{\mathbb{C}}
\newcommand{\ph}{\varphi}

\def\beal{\begin{align}}
\def\eeal{\end{align}}

\begin{document}

\title{{\sc{Pyessence}} - Generalised Coupled Quintessence Linear Perturbation Python Code - User Guide
\thanks{Supervisor, Dr Karim Malik and the Astronomy Unit of the School of Physics and Astronomy, Queen Mary, University of London}}
\author{Alexander Leithes}
\affiliation{Astronomy Unit, School of Physics and Astronomy,
Queen Mary University of London,Mile End Road, London, E1 4NS, UK}


\date{\today}

\begin{abstract}
This paper is a guide to the installation and use of the Python package {\sc{Pyessence}}. {\sc{Pyessence}} is designed to evolve linear perturbations to Coupled Quintessence models with a arbitrary number of cold dark matter (CDM) fluids and dark energy (DE) scalar fields as dictated by a given model. The equations are sufficiently general to allow for more exotic dark matter with a non-zero equation of state. Several example uses are included in order to demonstrate typical functionality to the potential user. {\sc{Pyessence}} is released under an open source modified BSD license and is available on Bitbucket.
\end{abstract}

\maketitle




\section{Introduction}
\label{Intro}

{\sc{Pyessence}} is a Python code designed to allow the testing of linearly perturbed coupled quintessence models with multiple CDM fluid species and multiple DE scalar fields; assisted coupled quintessence (see e.g. Refs.~\cite{Amendola:2014kwa,Amendola:1999dr}). For a more general overview of cosmological perturbation theory applied to multiple fluids see e.g. Refs.~\cite{MW2008,Malik:2004tf}.\\
The code allows two main approaches to investigating the viability of such models. Firstly, the ``stability" of these perturbations may be investigated. Here we use the word ``stability" rather loosely to mean the perturbations might experience runaway growth or ``explode"; models in which the perturbations have runaway growth may be excluded. Secondly, the power spectra or growth functions may be calculated to compare with observations. They may either prove to be outside current observational bounds (see e.g. Ref.~\cite{Macaulay:2013swa}) or provide deviations from the standard $\Lambda$CDM model of cosmology which would be detectable in future surveys e.g. SKA~\cite{Raccanelli:2015qqa} or Euclid~\cite{Kitching:2015fra}.\\
The code is designed to be as general as possible, with the form of the potential and other model specific parameters set in the MODEL.py module (while the equations within the code allow for more exotic forms of dark matter which might have non-zero equations of state i.e. a pressure, as in warm dark matter (WDM), see e.g. Ref.~\cite{Pagels:1981ke}). For any given model the code will either produce directly, or allow to be calculated, some quantities which may be compared with observations. Most directly the evolution of the density perturbations for the CDM species is produced by the code along with the background densities for the same. This in turn allows the power spectrum for the density perturbations to be generated by running the code for a range of wavenumbers, $k$. Furthermore, growth functions such as $g$, the evolution of the density contrast normalised by today's value i.e. $\frac{\delta}{\delta_0}$, or $f$, the e-fold derivative\footnote{here, e-fold is the logarithmic measure of time in terms of the expansion of the universe, such that $N = ln (\frac{a}{a_0})$, where $N$ is the number of e-folds, $a$ is the scale factor at a given time and $a_0$ is the scale factor today.} of the density contrast scaled to the density contrast i.e. $\frac{\delta'}{\delta}$ may be calculated from the data output and compared with, for example, $f\sigma_8$ measurements (see e.g. Ref.~\cite{Raccanelli:2015qqa}) or $fg$ see e.g. Refs.~\cite{Piloyan:2014gta,ACQ}. Finally, and related to this last point, although the code gives results in ``flat gauge" it is a simple matter to convert these into whichever gauge is required for a given research, or for comparison with existing literature e.g. the frequently used longitudinal or Newtonian gauge~\cite{Amendola:2014kwa,Raccanelli:2015qqa,Padmanabhan:2006kz}. We use perturbed flat (FLRW) cosmology with line element,
\be 
\label{IntDEds2}
ds^2=-(1+2\phi)dt^2+2aB_{,i}dt dx^i
+a^2\left(1+2(E_{,i j} - \psi \delta_{i j})\right)dx^idx^j \,,
\ee
where $x^i$ are the spatial coordinates, $t$ is coordinate time, a ``comma" denotes coordinate derivative, $E_{,i j}$ and $\psi$ are spatial metric perturbations, $B_{,i}$ is the shift function and $\phi$ the lapse function. Here the gauge is unspecified; to fix the gauge to flat we take only the scalar parts of the perturbations and set ${\psi}=0$, ${E}=0$. For more details on the theory behind the code please refer to the companion paper Ref.~\cite{ACQ}.\\
An advantage of this code is that it is relatively small and therefore fast. It can generate sufficient observables to allow the ruling out of regions of parameter space, or potentially a given model entirely, before embarking on more detailed analysis using larger codes with broader functionality e.g. CLASS~\cite{Blas:2011rf} or CAMB~\cite{Lewis:1999bs}.\\
The rest of this paper is set out as follows; Section \ref{Req} outlines the system requirements for the {\sc{Pyessence}} package. Section \ref{Install} contains installation instructions for the code. Section \ref{Modules} details the naming of the variables and other code specific features, as well as listing each of the modules; CONSTANTS.py, BACKGROUND.py, PERTURBED.py and MODEL.py. Finally, Section \ref{Examples} details some example applications of the {\sc{Pyessence}} code used in testing the code in development, as well to produce the first scientific results, as further insight into how to use the {\sc{Pyessence}} package.

\section{Requirements}
\label{Req}

{\sc{Pyessence}} was written and tested using Python 2.7.3 and should therefore work on higher versions. It may work with earlier versions but this has not been tested.\\
The core modules use Numpy, and these were developed and tested using v1.6.2.\\
The core modules use Scipy, and these were developed and tested using v0.10.1.\\
The various {\sc{Pyessence}} application examples e.g. EXAMPLE1.py, use Matplotlib to demonstrate plotting of results but this is not required for the core modules.

\section{Installation}
\label{Install}

Download the latest version of {\sc{Pyessence}} from either Bitbucket \cite{Bitbucket} or the {\sc{Pyessence}} website \cite{Pyweb}. If necessary unpack the zipped directories into a working directory. Make sure to add the paths for the Examples, General and Model folders to PYTHONPATH e.g. for a given session in Unix in bash shell type,
\begin{verbatim}
$export PYTHONPATH="$PYTHONPATH:/<directory_structure>/Examples"
$export PYTHONPATH="$PYTHONPATH:/<directory_structure>/General"
$export PYTHONPATH="$PYTHONPATH:/<directory_structure>/Model"
\end{verbatim}
each line followed by return. Adding the Data folder to PYTHONPATH is not necessary for the running of the core modules, however depending upon how the users application files are configured this might be necessary/advantageous. Once installed with the pathing updated it should be possible to run the application in the normal way from the command line e.g.
\begin{verbatim}
$python EXAMPLE1.py
\end{verbatim}
where in this example the output data file is saved in the Data folder.\\ Alternatively, to run the application a preferred interface may be used e.g. Spyder \cite{Spyder}. There is also a README.txt file included as a quick reference guide.

\section{Modules}
\label{Modules}

The variable and function labels are listed in a table in the README.txt file. The variables are stored in an array labeled $In[x]$, where $x$ runs from zero to $5+3A+4I$ where $I$ is the number of scalar fields and $A$ is the number of CDM fluids as defined by the dimensions of the couplings matrix.\\
The only module which the user would need to alter is the MODEL.py, which takes in the matrix for the couplings as an array labelled $C$ in the code, corresponding to $\C$ in the equations below. It is also where the value of $k$ is set, however, this may be overridden by the python module constructed to call the {\sc{Pyessence}} modules if, for example, stepping through $k$ space is required e.g. constructing power spectra. To explore the range of viable couplings for a given model, the $C$ matrix values may be overridden in a similar way. See section \ref{MODEL} for more details.

\subsection{CONSTANTS.py}
\label{CONSTANTS}

This module is the smallest and simply contains the constants used within the {\sc{Pyessence}} package. Any additional constants required if the code is modified should be put here. It contains the gravitational constant, $G$, 
\be
\label{kappa}
\kappa=(8 \pi G)^{\frac{1}{2}} , 
\ee
the Hubble parameter, $h$, Hubble's constant, $H_0$, and the critical density today, 

\be
\label{critdens}
\rho_{c(0)} (=\frac{3 H_0^2}{\kappa^2}). 
\ee

\subsection{BACKGROUND.py}
\label{BACKGROUND}

This module contains the background equations. In the first section are those which are not integrated, but are either constraints or useful quantities for the plotting of results. In the second section are the integrated equations. All the equations used here and in the PERTURBED.py module are taken from Ref.~\cite{ACQ}. A bar is used to denote background quantities. Please note; all the pressure terms in the equations throughout are replaced with appropriate equations of state terms within the code. The non-integrated equations are below.

\be
\label{FEIntDEBack}
H^2 = \frac{\kappa^2}{3} \left[\sum\limits_{\alpha} \bar{\rho}_{\alpha} + \sum\limits_{I} \frac{\dot{\bar{\ph}}_{I}^2}{2} + V \right] ,
\ee
is the Friedmann equation, for $\alpha$ fluids and $I$ scalar fields. $V$ is the potential for the scalar fields. The coordinate time derivative is denoted by dot. Derivatives with respect to fields are denoted by a ``comma".

\be
\label{FEdotIntDEBack}
\dot{H} = \frac{\kappa^2}{6 H} \left[\sum\limits_{\alpha} \dot{\bar{\rho}}_{\alpha} + \sum\limits_{I} (\dot{\bar{\ph}}_{I} {\ddot{\bar{\ph}}_{I}} + \dot{\bar{\ph}}_{I} V,_{\ph_{I}}) \right] ,
\ee
is the time derivative of $H$.

\be
\label{FieldDens}
\bar{\rho}_{\ph_{I}} = \frac{\dot{\bar{\ph}}_{I}^2}{2} + V  ,
\ee
are the energy densities of the scalar fields. The integrated equations are as follows.

\be
\label{dotrhorad}
\dot{\bar{\rho}}_r = - 4 H \bar{\rho}_r ,
\ee
is the evolution equation for the radiation energy density.

\be
\label{dotrhobar}
\dot{\bar{\rho}}_b = - 3 H \bar{\rho}_b ,
\ee
is the evolution equation for the baryon energy density.

\be
\label{dotrhoCDM}
\dot{\bar{\rho}}_{\alpha} = - 3 H (\bar{\rho}_{\alpha} - \bar{P}_{\alpha}) - \kappa \sum\limits_{I} \C_{I \alpha} (\bar{\rho}_{\alpha} - 3 \bar{P}_{\alpha}) \dot{\bar{\ph}}_{I} ,
\ee
are the evolution equations for the CDM energy densities, where $\C_{I \alpha}$ is the coupling matrix.\\
The next equations simply equate the functions for the time derivatives of the scalar fields, labelled $dx$ in the code, with the variables for the same, labelled $y$, within the code i.e. $\dot{\bar{\ph}}_{I}\equiv\dot{\bar{\ph}}_{I}$.

\be
\label{ddotph}
\ddot{\bar{\ph}}_{I} = - 3 H \dot{\bar{\ph}}_{I} - {V}_{, \ph_I} + \kappa \sum\limits_{\alpha} \C_{I \alpha} (\bar{\rho}_{\alpha} - 3 \bar{P}_{\alpha}) ,
\ee
are the second time derivatives of the scalar fields.

\subsection{PERTURBED.py}
\label{PERTURBED}

This module contains the perturbed equations. In the first section are those which are not integrated, but are either constraints or useful quantities for the plotting of results. In the second section are the integrated equations. A $\delta$ prefix is used to denote perturbed quantities. The non-integrated equations are below.

\be
\label{consPhi}
\phi = - \frac{\kappa^2}{2H} \left[\sum\limits_{\alpha} a \hat{v}_\alpha (\bar{\rho}_\alpha + \bar{P}_\alpha) - \sum\limits_{I} {\dot{\bar{\ph}}}_I \delta \ph_I \right] ,
\ee
is the constraint for the metric potential, $\phi$, where $\hat{v}_\alpha$ is $v_\alpha + B$, where $v_\alpha$ is the 3-velocity and $B$ is the shift.

\be
\label{consB}
B = \frac{3 \kappa^2 a}{2k^2} \left[ \frac{1}{3H} \left( \sum\limits_{\alpha} \delta \rho_\alpha - \sum\limits_{I} (\phi \dot{\bar{\ph}}^2_I - {\dot{\delta \ph}}_I {\dot{\bar{\ph}}}_I - V,_{\ph_I} \delta \ph_I ) \right) + \sum\limits_{I} {\dot{\bar{\ph}}}_I \delta \ph_I - \sum\limits_{\alpha} a \hat{v}_\alpha (\bar{\rho}_\alpha + \bar{P}_\alpha) \right] ,
\ee
is the constraint equation for $B$.

\be
\label{pertphdens}
\delta {\rho}_{\ph_{I}} = - \phi \dot{\bar{\ph}}^2_I + \dot{\bar{\ph}}_I {\dot{\delta \ph}}_I + V, _{\ph_I} \delta {{\ph}}_I ,
\ee
is the perturbed energy density for the scalar fields.

\be
\label{zeta}
\zeta = - \phi - H \left(\frac{\sum\limits_{\alpha} \delta \rho_\alpha}{\sum\limits_{\alpha} \dot{\rho}_\alpha} \right) ,
\ee
is the gauge invariant curvature perturbation, $\zeta$ (see e.g. Ref.~\cite{Carrilho:2015cma}). The integrated equations are as follows.

\be
\label{dotpertrhorad}
\dot{\delta \rho_r} = - 4 H \delta \rho_r + \frac{4 k^2}{3 a}(\hat{v}_r - B){\bar{\rho}}_r ,
\ee
is the evolution equation for the radiation perturbed energy density.

\be
\label{dotpertrhobar}
\dot{\delta \rho_b} = - 3 H \delta \rho_b + \frac{k^2}{a}(\hat{v}_b - B){\bar{\rho}}_b ,
\ee
is the evolution equation for the baryon perturbed energy density.

\be
\label{dotpertrhoCDM}
\dot{\delta \rho_\alpha} = - 3 H (\delta \rho_\alpha + \delta P_\alpha) + \frac{k^2}{a}(\hat{v}_\alpha - B)({\bar{\rho}}_\alpha + {\bar{P}}_\alpha) - \sum\limits_{I} \kappa \C_{I \alpha} ({\bar{\rho}}_\alpha - 3 {\bar{P}}_\alpha) {\dot{\delta \ph}}_I - \sum\limits_{I} \kappa \C_{I \alpha} (\delta \rho_\alpha - 3 \delta P_\alpha) {\dot{\bar{\ph}}}_I ,
\ee
are the evolution equations for the CDM perturbed energy densities.

\be
\label{dotvrad}
\dot{\hat{v}}_r = - \frac{\phi}{a} - \frac{\delta \rho_r}{4 a\bar{\rho}_r} ,
\ee
is the evolution equation for the 3-velocity for the radiation fluid.

\be
\label{dotvbar}
\dot{\hat{v}}_b = - H \hat{v}_b - \frac{\phi}{a} ,
\ee
is the evolution equation for the 3-velocity for the baryon fluid.

\be
\label{dotvCDM}
\dot{\hat{v}}_\alpha = \kappa \sum\limits_{I} \C_{I \alpha} (\bar{\rho}_\alpha - 3 \bar{P}_\alpha) \frac{\delta \ph_I}{a} + 3H \frac{\dot{\bar{P}}_\alpha}{\dot{\bar{\rho}}_\alpha} \hat{v}_\alpha - H\hat{v}_\alpha - \frac{\phi}{a} - \frac{\delta P_\alpha}{a({\bar{\rho}_\alpha} + \bar{P}_\alpha)} ,
\ee
are the evolution equations for the 3-velocities for the CDM fluids. Similarly to the BACKGROUND.py module, the next equation simply equates the function for the time derivative of the perturbed scalar fields, labelled $dpx$ within the code, with the variable for the same, labelled $py$, within the code i.e. ${\dot{\delta \ph}}_I\equiv {\dot{\delta \ph}}_I$.

\bea
\label{ddotdelph}
&{\ddot{\delta \ph}}_I& =  \left[ \frac{\kappa^2}{2H} \left( \sum\limits_{\alpha} \delta P_\alpha - \sum\limits_{I} (\phi \dot{\bar{\ph}}^2_I - {\dot{\delta \ph}}_I {\dot{\bar{\ph}}}_I + V,_{\ph_I} \delta \ph_I ) \right) - \frac{(3 H^2 + 2 \dot{H})}{H} \phi \right] {\dot{\bar{\ph}}}_I - \sum\limits_{J} V,_{\ph_I \ph_J} \delta \ph_J\\ \nonumber &-& \frac{k^2}{a^2} \delta \ph_I - \frac{k^2 B}{a} {\dot{\bar{\ph}}}_I - 2 V,_{\ph_I} \phi + 2 \sum\limits_{\alpha} \kappa \C_{I \alpha} ({\bar{\rho}}_\alpha - 3 {\bar{P}}_\alpha) \phi + \sum\limits_{\alpha} \kappa \C_{I \alpha} (\delta \rho_\alpha - 3 \delta P_\alpha) - 3 H {\dot{\delta \ph}}_I ,
\eea
are the second time derivatives of the perturbed scalar fields. Finally, the last function included at the end of the module is the array of all functions passed to the integrator, both background and perturbed.

\subsection{MODEL.py}
\label{MODEL}

This module is where the model being studied is defined, along with the setting of the wavenumber, $k$, placed here for convenience in the code structure. If the user wishes to write a module calling \PY~this value may be overwritten locally, for example if the user wanted to loop through $k$ values, or when plotting functions from multiple saved data sets for different $k$ values. Many of the parameters in the MODEL.py file included are specific to the sum of exponentials potential e.g.

\be
\label{sumofexppot}
V(\ph_1...\ph_I) = M^4 \sum\limits_{I} e^{-\kappa \lambda_I \ph_I} ,
\ee
used to test the \PY~code and give the first scientific results. This is simply included as one example from the large array of possible potentials, and the example MODEL.py contains many parameters set specifically for this example potential, and which would be altered if using a different potential e.g. the derivatives of the potential. Below only general quantities will be discussed.\\
$C$ is the array of the couplings. This has been entered directly in the included MODEL.py but may be loaded from a Numpy save file created separately, and for larger models with many CDM fluids and many scalar fields this will be the practical method. The module uses the dimensions of this array to determine the number of CDM fluids, assigned to variable $A$, and number of scalar fields, assigned to variable $I$. It also uses these to initialise the array of all integrated variables, $In$, the initial condition array, $f\_0$, and the CDM fluids equations of state array, $w$.\\
Function $V$ is the potential for a given model. Function $VP$ is the array of  derivatives of the potential with respect to the scalar fields. Function $VPP$ is the array of second derivatives of the potential with respect to the scalar fields. Those included in the MODEL.py file are for the sum of exponentials potential and have been entered manually as for a two CDM fluid, two scalar field model. More generally, for many CDM fluids and many scalar fields $VP$ and $VPP$ should be loaded from externally created saved arrays as per the $C$ array.

\section{Examples of Use}
\label{Examples}

These example Python files were created during the testing and initial use of the \PY~code. They are included in the package to give some guidance as to how the code may utilised, but are not intended to be prescriptive in its use.

\subsection{Example 1 - Matter and Radiation only Universe}
\label{Ex 1}

\begin{figure}[H]
\centerline{\includegraphics[angle=0,width=80mm]{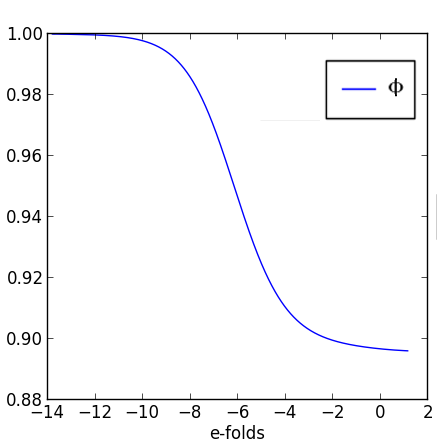}} 
\caption{Evolution of metric potential, $\phi$, for $k=0.01k_{eq}$, where $k_{eq}$ is the wavenumber for the horizon size at the time of matter-radiation equality.}
\label{padphi}
\end{figure}

This file was designed to evolve perturbations to just matter and radiation, with no CDM fluids or dark energy. This was compared to the same results in \cite{Padmanabhan:2006kz}. The file is included as EXAMPLEPAD.py. The corresponding model file is included as MODELPAD.py.\\
The imports section heads the file. Next, the range of times ($t\_i$ is the initial time, $t\_f$ the final time) and the stepping ($step$) is defined, in e-folds. After these are the initial conditions. In this example some of the perturbed initial conditions depended upon functions of the background, hence the split in the setting of the initial conditions seen in the module. For convenience an array of all times is created, $t\_out$.\\
Next the integrator is set up, with certain non-integrated useful functions calculated as the code runs. The $dopri5$ integration method is being used in this example but other possible integration methods are shown ``hashed out" within the code.\\
After the integrator has finished the results are saved ($fulloutput$), along with the time array ($t\_out$).\\
The next section demonstrates the plotting of some useful and interesting quantities either plotted directly from the results or calculated using them.\\
The first two sections plot the background and perturbed energy density for matter and radiation. The next two sections plot the density contrasts in the flat gauge in which the code is written and then plots the density contrasts converted to longitudinal gauge as in \cite{Padmanabhan:2006kz}. The next two sections plot longitudinal $\sigma$ (where $\sigma = \frac{3}{4} \delta_r - \delta_m$).\\
Next is flat $\sigma$. This is followed by the comoving curvature perturbation, $\zeta$, and then the metric potential, $\phi$, in flat gauge, the shift, $B$, and then $\phi$ in longitudinal gauge shown in Figure~\ref{padphi} for comparison with \cite{Padmanabhan:2006kz}.\\
Finally the 3-velocities for matter and radiation are plotted.

\subsection{Example 2 - $\Lambda$CDM}
\label{Ex 2}

\begin{figure}[H]
\centerline{\includegraphics[angle=0,width=100mm]{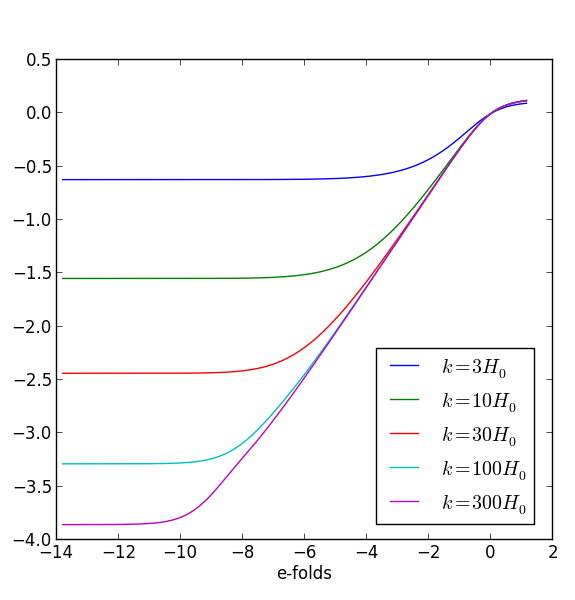}} 
\caption{Log of growth factor, g, $\frac{\delta}{\delta_0}$, subhorizon scales. $H_0$ is the Hubble constant.}
\label{gLCDMlong}
\end{figure}

The file for this example is included as LCDM.py. The model file corresponding to this is included as MODELLCDM.py. The layout is much as for Section \ref{Ex 1} with the following exceptions. This code was adapted from a test for one scalar field interacting with one CDM fluid. Standard $\Lambda$CDM behaviour was then achieved by flattening the potential and setting it to the same energy density as a cosmological constant today. The initial conditions for the scalar field and the scalar field velocity are then set to zero as are the field perturbation and field perturbation velocity. A small additional Python code called gANDfPLOTTER(long)LCDM.py, held in the $Data$ folder, was used to produce plots of the growth functions in longitudinal gauge over a range of $k$s. Figure~\ref{gLCDMlong} is included as an example output for log of the growth factor, $g$ ($=\frac{\delta}{\delta_0}$).
For further results and details see \cite{ACQ}.

\subsection{Example 3 - Assisted Coupled Quintessence - Transient Matter Domination}
\label{Ex 3}

\begin{figure}[H]
\centerline{\includegraphics[angle=0,width=100mm]{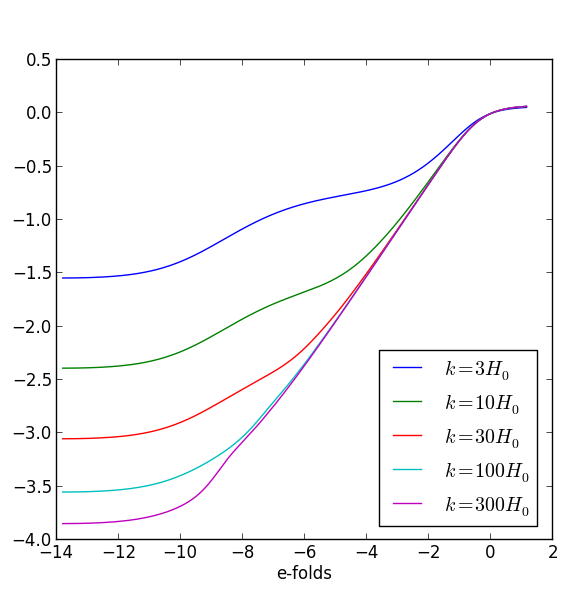}} 
\caption{Log of growth factor, g, subhorizon scales. $H_0$ is the Hubble constant.}
\label{gCQMDlong}
\end{figure}

The file for this example is included as EXAMPLE1.py. The model file corresponding to this is included as MODEL.py. The layout is much as for Section \ref{Ex 1} with the following exceptions. This code is for two scalar fields interacting with two CDM fluids. After initial radiation domination, an epoch of matter domination is entered which finally transitions to one of dark energy domination. A small additional Python code called gANDfPLOTTER(long).py, held in the $Data$ folder, was used to produce plots of the growth functions in longitudinal gauge over a range of $k$s. Figure~\ref{gCQMDlong} is included as an example output for log of the growth factor, $g$ ($\frac{\delta}{\delta_0}$).
For further results and details see \cite{ACQ}.

\section{Conclusion}
\label{Conclusion}

As detailed above, \PY~is designed to be a fast code for quickly performing initial testing of coupled quintessence models by constraining the allowed parameter space, or possible elimination or validation of models through comparison with observations. \PY~is released under an open source BSD license which can be found in the LICENSE.txt file included with this distribution. If \PY~is used in any published work the authors are kindly asked to cite this work and the related first implementation paper \cite{ACQ}.

\begin{acknowledgments}
The author is grateful to the collaborators on the main results paper \cite{ACQ}, Nelson Nunes, David Mulryne and Karim Malik, and also to Ian Huston, Pedro Carrilho and Phil Bull for useful and insightful discussions. Also, Tom Charnock for tea. AL is funded by an STFC studentship. The computer algebra package
{\sc{Cadabra}}\cite{Cadabra} was used in the derivation of some of the
equations.
\end{acknowledgments}



\end{document}